\documentclass[twocolumn,aps,reprint,superscriptaddress]{revtex4-2}
\usepackage{amsmath}
\usepackage[utf8]{inputenc}
\usepackage{txfonts}
\usepackage{amssymb}
\usepackage{gensymb}
\usepackage{graphicx}
\usepackage[a-2u]{pdfx}
\usepackage{xcolor}
\usepackage{comment}
\usepackage{booktabs}

\usepackage{hyperref}
\makeatletter

\renewcommand{\selectlanguage}[1]{}

\definecolor{lb}{rgb}{0.000, 0.400, 1.000}

\makeatother

\begin{document}

\title{Decoding the Micromagnetic Hamiltonian from Magnetic Fingerprints}

\author{Bradley J. Fugetta}
\affiliation{Department of Physics, Georgetown University, Washington, D.C. 20057,
USA}
\author{Anqi Liu}
\affiliation{Department of Computer Science, The Johns Hopkins University, Baltimore, Maryland 21218, USA}
\author{Kai Liu}
\affiliation{Department of Physics, Georgetown University, Washington, D.C. 20057,
USA}
\author{Amy Y. Liu}
\affiliation{Department of Physics, Georgetown University, Washington, D.C. 20057,
USA}
\author{Gen Yin}
\thanks{gen.yin@georgetown.edu}
\affiliation{Department of Physics, Georgetown University, Washington, D.C. 20057,
USA}

\begin{abstract}
Extracting intrinsic magnetic Hamiltonians directly from magnetometry is challenging due to the high dimensionality of the parameter space and the degeneracy induced by ensemble averaging. 
Here, we introduce a collection of deep convolutional neural networks (CNNs) to extract the full phenomenological micromagnetic Hamiltonian directly from the magnetic fingerprints encoded within First-Order Reversal Curves (FORCs). 
We validate this approach via closed-loop verification, re-creating the input magnetometry for both simulated and experimental FORCs. 
To mitigate false positives, we deploy an `Alice--Bob' parallel network that quantifies prediction uncertainty based on solely the information in FORCs without any additional ground-truth knowledge. 
This framework provides a robust, machine-learning-assisted approach to unravel the underlying spin behaviors in complex magnetic systems
\end{abstract}

\maketitle

\section{Introduction}

A central challenge in the study of magnetism and its diverse applications is quantitatively linking macroscopic magnetic behaviors to their underlying physical interactions. 
While magnetism is inherently a quantum phenomenon, classical micromagnetic Hamiltonians have demonstrated success in capturing key magnetic behaviors\cite{landau_theory_1992,brown_micromagnetics_1959}, including magnetization dynamics, spin waves, magnetic switching, and the behaviors of complex spin textures such as bubbles, vortices, and magnetic skyrmions\cite{abert_micromagnetics_2019,nagaosa_topological_2013,finocchio_magnetic_2016}.
However, the predictive power of these micromagnetic models relies on quantitatively knowing the complete set of parameters within the Hamiltonian, including the Heisenberg exchange, magneto-crystalline anisotropy, magnetic dipolar interaction, and the Dzyaloshinskii-Moriya interaction (DMI), among others\cite{vansteenkiste_design_2014,abert_neuralmag_2025}. 
In addition to these intrinsic material properties, extrinsic details such as microsctuctures, defects and impurities can also strongly affect magnetic behavior\cite{kronmuller_micromagnetism_1996,zhu_micromagnetic_1988}.
The experimental, quantitative extraction of these parameters is challenging, typically requiring a series of comprehensive, resource-intensive techniques coupled with iterative theoretical modeling\cite{woo_observation_2016,soucaille_probing_2016,burkert_giant_2004}. 
Certain interactions, such as the DMI, are particularly challenging to isolate and require specialized equipment and measurements to determine\cite{bode_chiral_2007,ferriani_atomic-scale_2008,heide_dzyaloshinskii-moriya_2008,torrejon_interface_2014,zakeri_asymmetric_2010,ma_dzyaloshinskii-moriya_2017,soucaille_probing_2016,kuepferling_measuring_2023,hrabec_measuring_2014,pinna_reservoir_2020}.
It is therefore highly desirable to develop methodologies capable of extracting the complete set of Hamiltonian parameters directly from readily accessible macroscopic measurements, such as spatial- and thermal-averaged magnetometry.
However, unlike the straightforward simulation of magnetometry from a known Hamiltonian, the inverse problem of extracting the Hamiltonian from magnetometry is notoriously difficult.
Since magnetometry measures the spatial average of all magnetic moments in a sample, it inevitably obscures the microscopic details of the spin textures, resulting in a highly degenerate parameter space.
Furthermore, inherent sample disorder and morphological variations drastically alter magnetic switching paths even in the adiabatic limit, further increasing the complexity.
Here we present a machine-learning driven solution to this complex inverse problem. 
In previous work, we demonstrated that DMI could be extracted from magnetometry data formatted as First-Order Reversal Curves (FORCs)\cite{pike_characterizing_1999,mayergoyz_mathematical_1991,davies_magnetization_2004, dumas_magnetic_2007, gilbert_quantitative_2014},  suggesting that the signatures of underlying spin-spin interactions are not completely obscured by ensemble averaging\cite{fugetta_machine-learning_2023}. 
In fact, analysis based on the distribution of FORCs has demonstrated its success in many complex magnetic systems\cite{davies_magnetization_2004,kirby_direct_2009,dumas_magnetic_2007,rahman_controlling_2009,gilbert_quantitative_2014,gilbert_realization_2015,burks_3d_2021}. 
Here we elevate our approach to a new scale, capable of extracting the full magnetic Hamiltonian directly from FORCs. 
This approach establishes a closed-loop verification, where the network's outputs are forward-simulated to confirm their agreement with the magnetometry inputs. 
Additionally, we introduce an `Alice-Bob' parallel network capable of assessing the confidence of the predictions based solely on the intrinsic features of the FORC data, without requiring any ground-truth knowledge. 
Our trained models are also validated by reproducing experimental FORCs using the predicted Hamiltonian parameters. 
This framework offers a reliable, data-driven workflow, providing significant leverage over traditional trial-and-error approaches when modeling various magnetic systems.

\section{Methods and Results}

\subsection{FORCs and Their Simulation}

The FORCs used in this work are obtained by micromagnetic simulations implemented by mumax$^{3}$\citep{vansteenkiste_design_2014}.
Due to the non-ergodic nature of classical spin models, the simulation time to arrive at the ground state of the spin texture is usually not reasonable for high-throughput data generation.  
Since the materials of our interest usually have high Curie temperatures ($T_C\gg300\thinspace\textrm{K}$), zero-temperature dynamics is enough to capture the spin-texture evolution measured at room temperature. 
This removes the thermalization term in the effective field, avoiding the known scaling artifacts associated with finite-temperature micromagnetics \citep{grinstein_coarse_2003,hahn_temperature_2019}.
%
%
To precisely capture the FORCs with fast simulations, we need to identify the maximum time step $\Delta t$ that does not modify the instantaneous susceptibility $\frac{dm_z(t)}{dH_z(t)}$. 
Here $m_z=\frac{M_z}{M_S}$ is the normalized perpendicular magnetization and $H_z(t)$ is a time-dependent applied magnetic field normal to the thin film. 
The slew rate $\frac{dH_z}{dt}$ is adaptively chosen based on the maximum torque ($\tau_z$) to reduce the simulation time. 
To estimate $\tau_{z,\max}$ for each simulation, we use the Landau-Lifshitz-Gilbert (LLG) equation under the assumption that $\mathbf{H}_\mathrm{eff}$ lies predominantly along the $z$-axis on average, yielding
\begin{equation}
    \tau_{z}
= -\gamma_\textrm{LL}\,\frac{\alpha'}{1+\alpha'^2}\,
\mu_0 H_{\mathrm{eff},z}\,(1 - m_z^2).
\label{eq:llg_spinflip}
\end{equation}
This assumes a single, averaged classical magnetization $\mathbf{m}(t)=\frac{ \mathbf{M}(t) }{M_S}$, where $\gamma_\textrm{LL}$ is the Landau-Lifshitz gyromagnetic ratio and $\mathbf M(t)$ denotes position-averaged magnetization at time $t$. 
Since the magnetization is an averaged value, we replace the Gilbert damping factor $\alpha$ with an effective value $\alpha'$. 
Specifically, we first saturate the simulation at $                                          -1\thinspace\textrm{T}$, instantaneously apply a constant field of $+1\thinspace\textrm{T}$, and then take $\tau_{z,\textrm{max}}=\tau_z(m_z = 0)$. 
The value of $\alpha'$ is then determined by solving Eq.~\ref{eq:llg_spinflip} at $m_z=0$. 
\begin{figure}
\begin{centering}
\includegraphics[width=\columnwidth]{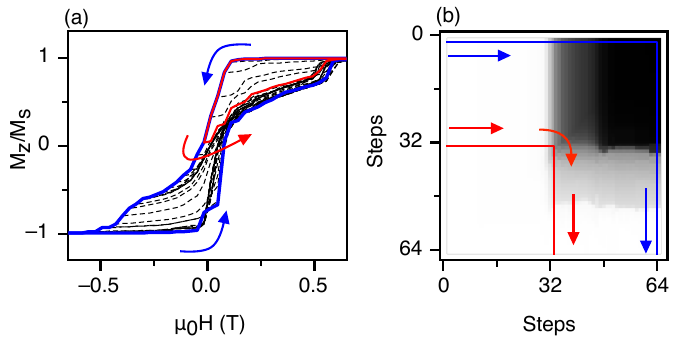}
\par\end{centering}
\caption{\textbf{FORCs simulation and 2D rearrangement.}
(a) A typical family of FORCs in our dataset.
One minor loop is highlighted by the red curve. The full hysteresis loop is denoted by the solid blue curve. The FORCs are illustrated by the dark dashed lines.
(b) The rearrangement of FORCs into a 2D image, with each pixel denoting one discrete point during the field scan.
The white-dark color scale denotes the normalized magnetization quantized to a single-byte integer.
\label{fig:FORCsAndTheRearrangement2D}}
\end{figure}

\begin{figure*}
\begin{centering}
\includegraphics[width=\textwidth]{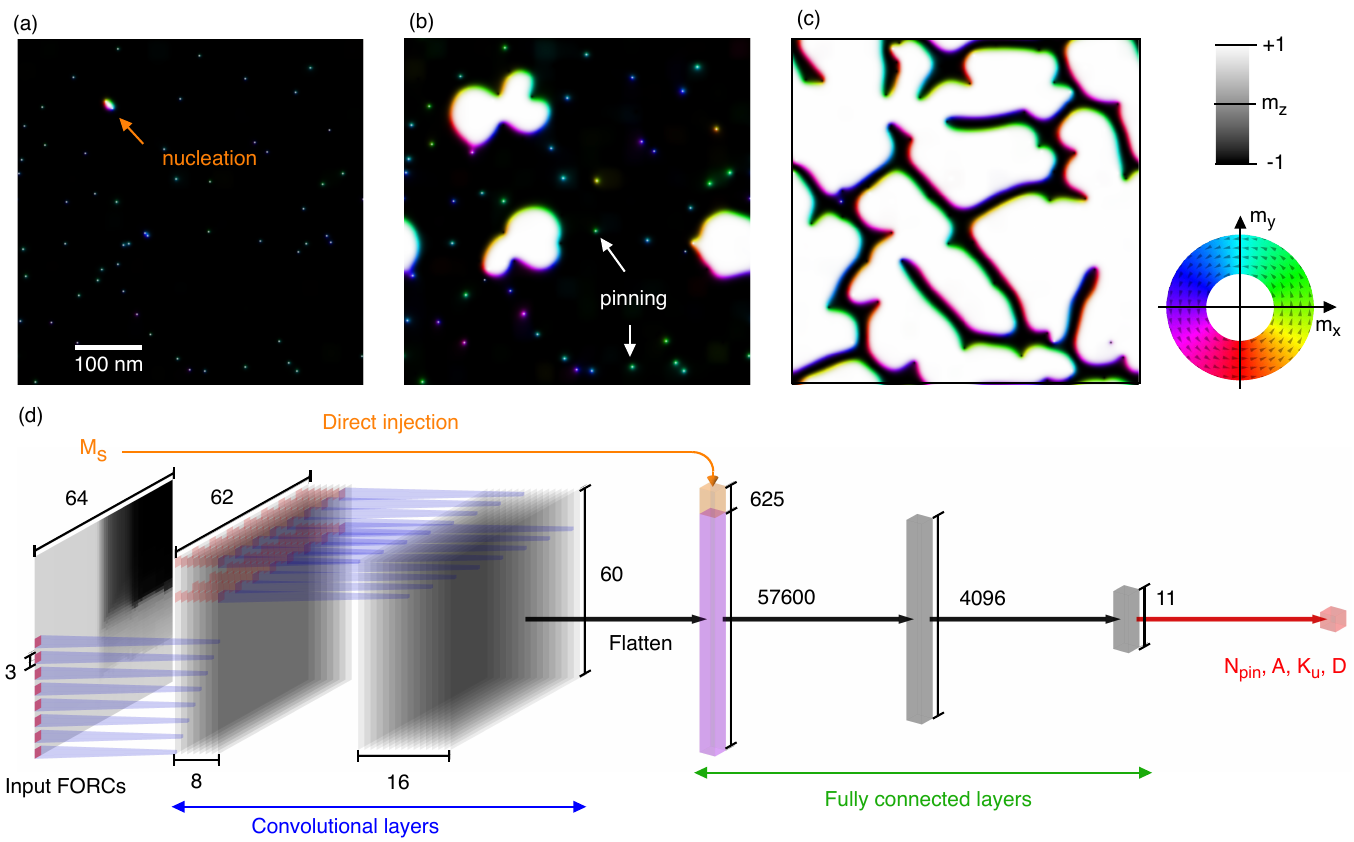}
\par\end{centering}
\caption{\textbf{Data generation and the CNN structure.}
(a) One example of the beginning stage of a FORC. The orange arrow denotes the nucleation of a magnetic domain. 
The faint colored dots correspond to the pinning sites stuck with positive magnetization.
(b) The development of domains propagating through pinning sites. The local azimuthal angle of the magnetization is denoted by the color wheel. 
(c) The final stage of magnetic reversal after (a) and (b). 
(d) The structure of our predicting CNNs. Each micromagnetic Hamiltonian parameter is predicted by its own CNN trained independently.   
\label{fig:DomainDevelopmentAndCNN}}
\end{figure*}

When scanning $H_z(t)$, we periodically examine and update the slew rate $\frac{\partial H_z}{\partial t}$ based on the current and recent state of the system.
Each update is performed if $\mu_0 H_z$ changes by $0.02~\textrm{T}$ or $m_z$ changes by 0.05, whichever comes first. 
Each update period is roughly $10^3$ times larger than the integrator time step. 
At the end of each update period, the present point on the hysteresis curve $(H_z, M_z)$ is recorded together with the two previous periods. 
We then estimate the differential susceptibility
$\chi_{zz} = \frac{\partial M_z}{\partial H_z}$. 
This is then combined with $\tau_{z,\textrm{max}}$, providing a safe slew rate for the next section:
$
\frac{\partial H_z}{\partial t}
= \frac{\partial m_z}{\partial t}\frac{\partial H_z}{\partial m_z}
= \frac{\tau_{z,\max}}{\chi_{zz}}.
$
In practice, we used an extra safety factor $\frac{\partial H_z}{\partial t}\rightarrow0.05\frac{\partial H_z}{\partial t}$ to make sure the scan is within a reasonable adiabatic limit. 
We also imposed hard upper and lower bounds on the slew rate to avoid numerical noise in the derivative.
One example of the simulated FORCs is illustrated in Fig.~\ref{fig:FORCsAndTheRearrangement2D}(a). 
The simulation result is then discretized and re-arranged into a $64\times64$ image as the standard input format of our predicting CNNs, as shown in Fig.~\ref{fig:FORCsAndTheRearrangement2D}(b).

\subsection{Generating the Dataset}

Our dataset was constructed from micromagnetic simulations using a $256 \times 256 \times 2$ spin lattice with a cell size of $2\thinspace\mathrm{nm} \times 2\thinspace\mathrm{nm} \times 4\thinspace\mathrm{nm}$. 
The base material parameters defining the Hamiltonian, including the saturation magnetization ($M_S$), the exchange stiffness ($A_\mathrm{ex}$), the uniaxial anisotropy energy density ($K_u$), and the DMI ($D$), were sampled to broadly cover the phase space of technologically relevant magnetic materials, informed by the open-access Novomag database \citep{sakurai_discovering_2020,balasubramanian_synergistic_2020}. 
Specifically, $M_S$ was drawn uniformly from $200$ to $1400\thinspace\mathrm{kA/m}$, and $\ln(K_u)$ was drawn from a triangular distribution peaking at its maximum of $\ln(1.5 \times 10^6\thinspace\mathrm{J/m^3})$ with a minimum of $\ln(10^3\thinspace\mathrm{J/m^3})$. 
To mimic large thin films, we imposed a periodic boundary condition with $256$ repetitions in the $x\textrm{-}y$ plane. 
To compensate for the absence of thermal fluctuations, which naturally nucleate domains and impede domain wall motion, we introduced two explicit models of structural disorder. 
First, we incorporated $N_\mathrm{pin}$ localized defects, chosen uniformly from $0 \leq N_\mathrm{pin} \leq 255$. 
Each pinning site was modeled as a single cell with a randomized initial spin orientation and an artificially high out-of-plane anisotropy of $K_u = 50\thinspace\mathrm{MJ/m^3}$. 
These sites act as nucleation centers and domain wall pinning centers, replicating the stochastic, jerky reversal dynamics observed in experimental FORC diagrams \citep{jeudy_pinning_2018}. 
Several different stages of an example domain evolution are shown in Figs.~\ref{fig:DomainDevelopmentAndCNN}(a-c). 
To emulate the inherent randomness of a polycrystalline sample, the grid was further partitioned into $255$ distinct regions where the anisotropy magnitude followed $K_u = K_{u,\mathrm{sample}}(1+\Delta)$. 
Here, $\Delta$ is drawn from a normal distribution $\mathcal{N}(0, \sigma_K)$ with a mean of zero and standard deviation $\sigma_K$. 
Similarly, the easy-axis orientation $\hat{n}$ was dispersed by assigning each region a random azimuthal angle $\phi \in [0, 2\pi]$ and a polar angle $\theta$ drawn from a normal distribution $\mathcal{N}(0, \sigma_\theta)$.
Because $A_\mathrm{ex}$ and $M_S$ jointly determine the exchange length, $l_\mathrm{ex} = \sqrt{2 A_\mathrm{ex} / \mu_0 M_S^2}$, we employed a constrained sampling algorithm to ensure numerical stability and guarantee the formation of multi-domain states within our $512\thinspace\mathrm{nm}$ simulation box. 
We required $l_\mathrm{ex} > 5\thinspace\mathrm{nm}$ and targeted a mean $l_\mathrm{ex}$ of $10\thinspace\mathrm{nm}$. 
In this procedure, $M_S$ was first drawn from its uniform distribution and a target $l_\mathrm{ex}$ was drawn from a normal distribution $\mathcal{N}(10\thinspace\mathrm{nm}, 1\thinspace\mathrm{nm})$. 
Within these constraints, $A_\mathrm{ex}$ was then calculated and fed to the simulation. 
If the resulting $A_\mathrm{ex}$ fell outside the physical bounds of $1$ to $35\thinspace\mathrm{pJ/m}$, it was instead drawn uniformly within those bounds and then $l_\mathrm{ex}$ was recalculated to verify the $5\thinspace\mathrm{nm}$ minimum threshold.
To maximize the density of high-quality data and avoid wasting computational resources, each initialized sample underwent preliminary screening. 
Samples were required to achieve full saturation at $+1.0\thinspace\mathrm{T}$, and demonstrate sufficient hysteretic area evaluated by simulating half of the major loop. 
Only those parameter sets meeting these criteria (success rate of $\sim 20\%$) were fully simulated to produce the final dataset. 
The absolute bounds for all sampled parameters are summarized in Table~\ref{table:theBounds}.

\begin{table}[h]
\caption{\textbf{The bounds of simulation parameters}\label{table:theBounds}}
\centering
\begin{tabular}{lccccccc}
\toprule 
 & $\mathrm{M_{s}}$ & $\mathrm{N_{pin}}$ & $\mathrm{A_{ex}}$ & $\mathrm{K_u}$ & $\mathrm{D}$ & $\mathrm{\sigma_{\theta}}$ & $\mathrm{\sigma_{K}}$\tabularnewline
\, & A/m & $\#$ & J/m & J/m$^{3}$ & J/m$^{2}$ & $^{\circ}$ & $\%$\tabularnewline
\midrule 
Min & $\,\,\,\,2\times10^{5}$ & $\,\,\,\,\,\,0$ & $\,\,\,1\times10^{-12}$ & 1000 & 0 & 0 & 0\tabularnewline
Max & $14\times10^{5}$ & 255 & $35\times10^{-12}$ & $1.5\times10^{6}$ & 0.005 & 10 & 20\tabularnewline
\bottomrule
\end{tabular}
\end{table}

\subsection{Training the CNNs}

We implemented and trained convolutional neural networks (CNNs) \cite{kawaguchi_determination_2021,singh_application_2019,kwon_magnetic_2020,wang_machine_2020} using the PyTorch machine learning framework. 
Our CNN contains two convolutional layers followed by two fully connected (dense) layers. 
The structure is illustrated in Fig.~\ref{fig:DomainDevelopmentAndCNN}(d).
The input FORC images are processed by the first convolutional layer to produce 8 feature maps with spatial dimensions of $62 \times 62$. 
A subsequent convolutional layer extracts higher-order spatial features, resulting in 16 feature maps of $60 \times 60$. 
These feature maps are then flattened into a one-dimensional tensor of $57,600$ nodes.
Crucially, because $M_S$ is a macroscopic quantity readily accessible via magnetometry, we explicitly inject it into the network to constrain the parameter search space. 
We concatenate the flattened convolutional features with an array of $625$ nodes carrying the normalized $M_S$ value, yielding a combined input of $58,225$ nodes for the dense block. 
This concatenated array is then fed into the first fully connected layer, which compresses the representation down to 4,096 nodes. 
These latent activations are then passed to the 11 nodes in the final fully connected layer, from which a single prediction of one Hamiltonian parameter is made using a `distribution matching' method\cite{fugetta_machine-learning_2023}. 
A dropout probability of $0.5$ was applied to all layers except the first convolutional layer.
A series of CNNs of the same structure were constructed and trained independently to predict each Hamiltonian parameter. 
The network weights and the CNN kernels were optimized using the ADADELTA adaptive learning rate algorithm \citep{zeiler_adadelta_2012}. 
We used a base learning rate of $\gamma = 0.01$, a momentum of $\rho = 0.9$, and weight decay of $\lambda = 0$. 
The generated dataset was partitioned into an $80/20$ train-validation split. 
The $20\%$ validation set was strictly withheld from the gradient updates and used exclusively to evaluate the performance and monitor for overfitting at the conclusion of each training epoch.

\section{Discussion}
%

Predictions for $N_{\mathrm{pin}}$, $A_{\mathrm{ex}}$, $K_{u}$, and $D$ given by our trained CNNs significantly outperformed random guessing. 
%
%
%
%
\begin{figure}
\begin{centering}
\includegraphics[width=1\columnwidth]{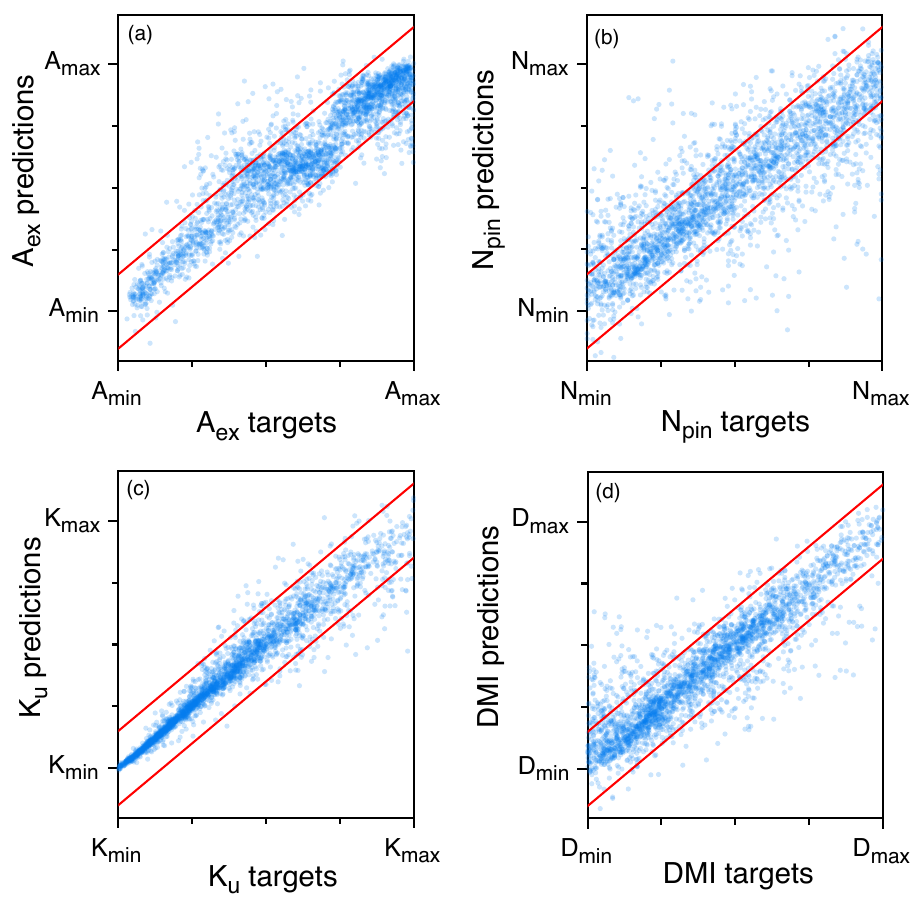}
\par\end{centering}
\caption{\textbf{CNN Predictive Performance among the validation set.} The four panels illustrate the predicted values versus the targets: (a) $A_{\mathrm{ex}}$, (b) $N_{\mathrm{pin}}$, (c) $K_{u}$, and (d) $\textrm{DMI}$. Red lines mark a 15\% error threshold. Each point is illustrated with $20\%$ opacity. 
\label{fig:ParameterTestingResults}}
\end{figure}
The validation results are illustrated in Fig.~\ref{fig:ParameterTestingResults}(a-d), where the horizontal axes correspond to the target values $x'_i$, whereas the vertical ones represent the predicted values $x_i$. 
Although errors do exist in all four cases, the vast majority of predictions successfully fall within our error threshold of 0.15 (red lines) of the full range.
Quantitatively, we can measure the accuracy using the mean absolute error, $\langle|\Delta x|\rangle=\langle|x_i-x'_i|\rangle$, and the Pearson correlation coefficient, $r$:
\begin{equation}
r = \frac{\sum_{i=1}^{n} (x_i - \langle x \rangle)(x'_i - \langle x' \rangle)}{\sqrt{\sum_{i=1}^{n} (x_i - \langle x \rangle)^2 \sum_{i=1}^{n} (x'_i - \langle x' \rangle)^2}}
\end{equation}
where $i$ runs through all the data points, and $\langle\cdots\rangle$ denotes the mean. 
Since the parameters are normalized to $[0,1]$, $\langle|\Delta x|\rangle$ represents the dimensionless fractional uncertainty. 
For these four parameters, the statistics are: $\langle|\Delta N_{\mathrm{pin}}|\rangle = 0.11$ ($r_{\mathrm{N}} = 0.84$), $\langle|\Delta A_{\mathrm{ex}}|\rangle = 0.07$ ($r_{\mathrm{A}} = 0.92$), $\langle|\Delta K_{u}|\rangle = 0.04$ ($r_{\mathrm{K}} = 0.97$), and $\langle|\Delta D|\rangle = 0.07$ ($r_{\mathrm{D}} = 0.90$), respectively.

\subsection{Simulation Validation}

\begin{figure*}
\begin{centering}
\includegraphics[width=\textwidth]{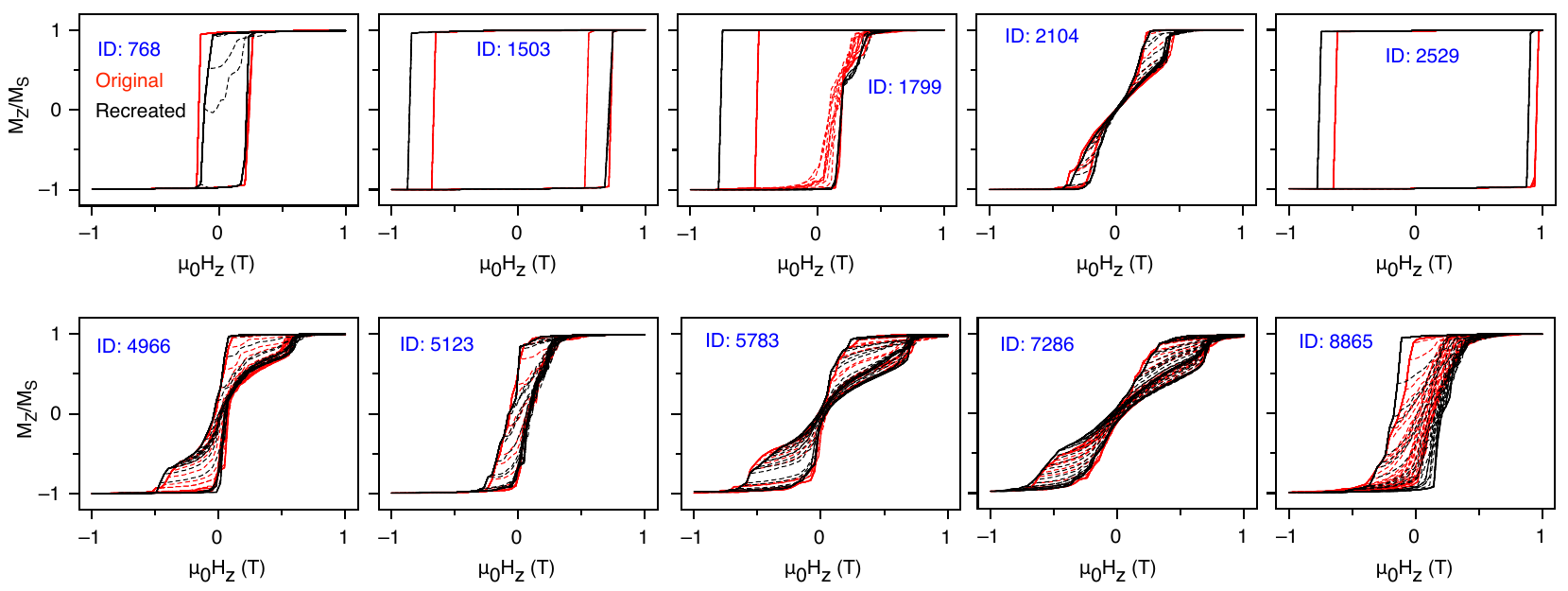}
\par\end{centering}
\caption{\textbf{Close-loop validation using simulations. }
Ten randomly chosen samples $\vec y$ from the original validation set (red) are compared to the FORCs recreated by new simulations using the predicted micromagnetic Hamiltonians $\mathcal{\vec{F}}_{\mathrm{sim}}(\mathcal{\vec{F}}_{\mathrm{CNN}}(\vec{y}))$.
\label{fig:SimulationValidation}}
\end{figure*}

Relying exclusively on scalar error metrics (e.g., $\langle|\Delta x|\rangle$ and $r$) to evaluate CNN performance implicitly assumes that all Hamiltonian parameters uniformly influence the magnetization reversal process. 
Physically, however, the details of the FORCs are governed by a complex, non-linear interplay of these parameters. 
For example, in a strongly exchange-coupled regime (large $A_{\mathrm{ex}}$), the macroscopic micromagnetic state becomes largely insensitive to the DMI ($D$). 
In such cases, a significant prediction error in $D$ may translate to a physically negligible difference in the resulting FORCs.
To robustly evaluate whether our CNNs have learned the underlying physics, we established a qualitative `closed-loop' validation metric. 
We define the micromagnetic simulation (via mumax$^3$) as the forward function $\mathcal{\vec{F}}_{\mathrm{sim}}(\vec{x}) = \vec{y}$, mapping the Hamiltonian parameters ($\vec{x}$) to a FORC image ($\vec{y}$). 
Training the CNN approximates the inverse: $\mathcal{\vec{F}}_{\mathrm{CNN}} \approx \mathcal{\vec{F}}_{\mathrm{sim}}^{-1}$. 
In the previous section we have evaluated the parameter-space error, $\langle |\Delta x|\rangle=\langle|\mathcal{\vec{F}}_{\mathrm{CNN}}(\mathcal{\vec{F}}_{\mathrm{sim}}(\vec{x})) - \vec{x}|\rangle$. 
Here, we further assess the fidelity of the observation-space by comparing the reconstructed output $\mathcal{\vec{F}}_{\mathrm{sim}}(\mathcal{\vec{F}}_{\mathrm{CNN}}(\vec{y}))$ directly against the ground-truth target $\vec{y}$.
For this analysis, ten samples were randomly selected from the validation set. 
We passed their simulated FORCs through the CNNs to extract the full suite of predictions ($N_{\mathrm{pin}}, A_{\mathrm{ex}}, K_{u}, D$). 
These predictions, along with the known $M_S$, were re-injected into mumax$^3$ to reconstruct the input FORCs. 
As illustrated in Fig.~\ref{fig:SimulationValidation}, reconstructed curves using the predicted Hamiltonian parameters successfully reproduce key magnetic signatures in most cases, including the coercive field distribution, the steepness of the reversal and the overall shapes of the minor loops. 
More significant disagreements can be seen in cases with featureless FORCs where almost all FORCs are coinciding with the major loop (IDs: 768, 1503 and 2529) due to the abrupt switching. 
This observation suggests that the Hamiltonian information is mainly hidden in those FORCs that are deviating from the major loop, forming feature-full minor loops. 
These FORCs are typically induced by the domain developments during switching, containing the information of both the intrinsic Hamiltonian and the extrinsic disorder. 
This successful closed-loop validation also highlights the robustness of our \textit{decoupled parameter prediction}. 
Because each Hamiltonian parameter is extracted by a strictly independent CNN, the networks cannot rely on internal cross-correlations to optimize their outputs. 
This further suggests that the fingerprints of all Hamiltonian terms are independently hidden in the FORCs.

\subsection{Uncertainty Quantification}
\label{subsec:uncertainty}

\begin{figure*}
\begin{centering}
\includegraphics[width=1\textwidth]{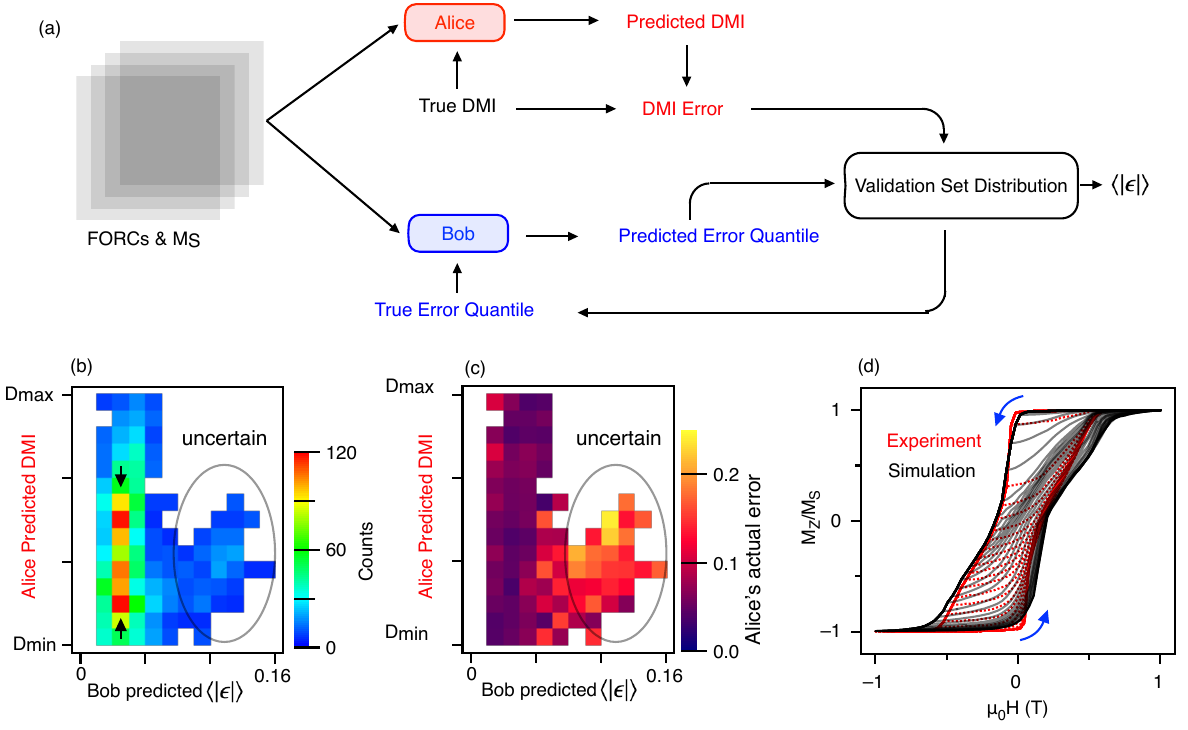}
\par\end{centering}
\caption{\textbf{Alice-Bob network and the experimental validation.} 
(a) The structure of the parallel workflow of the Alice-Bob network predicting the uncertainty without ground truth.
(b) The distribution of Bob's prediction within the full range of DMI values predicted by Alice. The blue-green-red color scale illustrates the count of points within each bin. 
(c) The average of the actual error of Alice illustrated for all the bins shown in (b). The color scale illustrates the normalized quantitative error between $0$ and $1$.  
(d) The experimentally measured FORCs (red) compared to the simulated FORCs (dark) using the predicted Hamiltonian parameters and the $M_S$.
\label{fig:ExperimentalValidationAndUncertaintyEstimation}}
\end{figure*}

In application scenarios, it is often useful to have reliable accuracy assessment for all Hamiltonian parameters immediately after the predictions are made. 
As demonstrated in Figs.~\ref{fig:SimulationValidation}, accurate predictions of the Hamiltonian typically require rich details in the FORCs.
This suggests that information hidden in the FORCs can be used as indicators of the prediction performance even without any ground-truth knowledge. 
It is thus compelling to establish another model to flag those FORCs that are particularly difficult for predicting CNNs to handle. 
To achieve this, we established a parallel dual-network architecture, which we refer to as the `Alice-Bob' network, as illustrated in Fig.~\ref{fig:ExperimentalValidationAndUncertaintyEstimation}a. 
The predicting network, `Alice', extracts the Hamiltonian parameters from the input FORCs and the corresponding $M_S$. 
Simultaneously, a secondary network, `Bob', is tasked with mapping the exact same input to Alice's prediction error.
Directly training Bob using Alice's quantitative errors is challenging, since most of the predictions made by Alice are accurate. 
To prevent Bob from mindlessly predicting low error, we trained Bob to predict the normalized quantile of Alice's errors rather than directly using the error values. 
This forces Bob to distinguish features across the entire error spectrum, instead of focusing on the quantitative error of each prediction. 
Once the normalized quantile of error is accurately captured by Bob, it is then straightforward to recover the quantitative uncertainty $\langle|\epsilon|\rangle$ using Alice's error distribution within the validation set. 
This workflow is illustrated in Fig.~\ref{fig:ExperimentalValidationAndUncertaintyEstimation}(a).
%
%

%
We first examine the distribution of Bob's error prediction $\langle|\epsilon|\rangle$ within the full parameter range. 
Taking DMI as an example, we show 
the distribution of the predicted continuous values of $D$ by Alice and the predicted error by Bob using $15$ discrete bins.
To ensure statistical clarity and suppress noise, bins containing fewer than five data points have been masked out. 
The statistics is illustrated in Fig.~\ref{fig:ExperimentalValidationAndUncertaintyEstimation}(b). 
According to Bob's prediction, the vast majority of Alice's predictions form a dense, vertical `main branch' centering at $\langle|\epsilon|\rangle<0.05$ (peaks denoted by dark arrows). 
A small number of DMI predictions near the mid-range values of $D$ are recognized to be inaccurate by Bob, forming a horizontal `uncertain branch' highlighted by the oval. 
However, the total number of samples falling into this region is relatively small, consistent with the good overall performance of Alice. 

To demonstrate Bob's performance, we check if those predictions of DMI flagged by Bob are indeed inaccurate. 
This is demonstrated in Fig.~\ref{fig:ExperimentalValidationAndUncertaintyEstimation}(c), where the color of each pixel is replaced by the average of the true error made by Alice within each bin. 
%
%
%
%
The horizontal color gradient demonstrates that Bob's predicted error $\langle|\epsilon|\rangle$ tightly correlates with Alice's true error across the entire parameter domain. 
The separation between the branches is clear: the heavily populated `accurate branch' reliably exhibits errors well below $0.1$, while the sparsely populated uncertain branch decisively isolates the high-error outliers.

\subsection{Experimental Validation}

Beyond the closed-loop validation within the micromagnetic theory, we further validated the CNN predictions using experimental data. 
We characterized a Co/Pd multilayer thin film with the stacking of Si/Pd(20)/[Co(0.4)/Pd(0.6)]$_{59}$/Co(0.4)/Pd(5). 
The numbers in parentheses denote the thickness in nanometers. 
This thin film is known to have a robust perpendicular magnetic anisotropy and a tunable hysteresis behavior modulated by the deposition pressure\citep{kirby_direct_2009,kirby_vertically_2010,greene_deposition_2014}. 
The magnetometry was performed in a Vibrating Sample Magnetometer (VSM), and the measured FORCs were discretized and then mapped into a CNN-compatible image.
Due to the setting of our modeling Hamiltonian and the design of our CNNs, an averaged $M_S$ is needed for this multilayer thin film containing both the magnetic and non-magnetic layers. 
Assuming sharp interfaces, the $M_S$ value corresponding to the total Co volume yields $M_S=1430\thinspace\textrm{kA/m}$, which is consistent with bulk Co. 
Normalizing the net magnetization by the total Co/Pd superlattice volume reduces this value to approximately $560\thinspace\textrm{kA/m}$. 
To account for this ambiguity, we swept $M_S$ across the range of $[470, 1400]\thinspace\textrm{kA/m}$, generating a distinct set of CNN parameter predictions for each $M_S$ input.
To suppress the extra degree of freedom introduced by the distribution of the pinning sites, we averaged the simulated FORCs over $20$ different profiles of pinning sites within the fixed value of $N_\textrm{pin}$ predicted by the CNN. 
The optimal agreement occurs at $M_S=1307\thinspace\textrm{kA/m}$, as illustrated in Fig.~\ref{fig:ExperimentalValidationAndUncertaintyEstimation}(d). 
While the reconstruction (red) is not a perfect match to the experimental measurement (dark), it captures the primary macroscopic magnetic signatures such as the coercivity, the susceptibility and the overall squareness of the full hysteresis loop. 
We note that our reconstruction of the FORCs in Fig.~\ref{fig:ExperimentalValidationAndUncertaintyEstimation} has lower agreement with the experimental input compared to the best-performing cases in the close-loop validations shown in Fig.~\ref{fig:SimulationValidation}. 
This suggests that the size of our training set and the complexity of the CNNs are already sufficient to capture the modeling Hamiltonian within its approximations. 
To further enhance this agreement, the Hamiltonian must include higher-order terms and intricate microstructural variations, suggesting rich research opportunities in the future. 

\section{Conclusion}

In conclusion, our work demonstrates that the spatial-averaged nature of magnetometry does not completely obscure the microscopic physics of the spin system. 
Because the minor hysteresis loops in FORCs contain rich structural features arising from domain wall pinning and the competition among distinct energy terms, it is possible to extract the full micromagnetic Hamiltonian including the exchange stiffness ($A$), the uniaxial anisotropy ($K_u$), the pinning center density ($N_\textrm{pin}$), and the Dzyaloshinskii-Moriya interaction ($D$). 
When attempting to model complex magnetic behaviors, these data-driven predictions offer a crucial advantage over traditional, time-consuming trial-and-error fitting methods. 
By extracting the full Hamiltonian, we can now `close the loop' by re-simulating the system using the CNN-predicted parameters to directly verify that they successfully reproduce the original magnetometry data. 
Our addition of experimental verification confirms that this framework can translate from simulated datasets to real-world physical systems.
Crucially, we have shown that the physical features within the magnetometry data itself encode intrinsic indicators of prediction reliability. 
Using our Alice-Bob network, we can systematically estimate the uncertainty of these predictions without reliance on ground-truth knowledge.
This proof-of-concept work strongly motivates the expansion toward more sophisticated, realistic micromagnetic models including both the intrinsic magnetic Hamiltonian and the extrinsic, microscopic morphological details.
This foundation is a critical step toward fully decoding the micromagnetic physics hidden within macroscopic magnetic fingerprints.

\begin{acknowledgments}
This work is supported in part by the National Science Foundation (US) under Grant \# DMR-2440337 and by the Defense Advanced Research Projects Agency (DARPA) under Contract No. D26AP50003.
The material does not necessarily reflect the position or the policy of the government and no official endorsement should be inferred.
The computations in this work are performed on Bridges-2 at Pittsburgh Supercomputing Center through allocation PHY230018 from the Advanced Cyber infrastructure Coordination Ecosystem Services \& Support (ACCESS) program, which is supported by National Science Foundation (US) grants \#2138259, \#2138286, \#2138307, \#2137603, and \#2138296.
\end{acknowledgments}
\bibliography{referencesV4.bib}

@article{abert_neuralmag_2025,
	title = {{NeuralMag}: an open-source nodal finite-difference code for inverse micromagnetics},
	volume = {11},
	copyright = {2025 The Author(s)},
	issn = {2057-3960},
	shorttitle = {{NeuralMag}},
	url = {https://www.nature.com/articles/s41524-025-01688-1},
	doi = {10.1038/s41524-025-01688-1},
	language = {en},
	number = {1},
	urldate = {2026-06-22},
	journal = {npj Computational Materials},
	publisher = {Nature Publishing Group},
	author = {Abert, C. and Bruckner, F. and Voronov, A. and Lang, M. and Pathak, S. A. and Holt, S. and Kraft, R. and Allayarov, R. and Flauger, P. and Koraltan, S. and Schrefl, T. and Chumak, A. and Fangohr, H. and Suess, D.},
	month = jun,
	year = {2025},
	pages = {193},
}

@article{pinna_reservoir_2020,
	title = {Reservoir {Computing} with {Random} {Skyrmion} {Textures}},
	volume = {14},
	url = {https://link.aps.org/doi/10.1103/PhysRevApplied.14.054020},
	doi = {10.1103/PhysRevApplied.14.054020},
	number = {5},
	urldate = {2026-06-22},
	journal = {Physical Review Applied},
	publisher = {American Physical Society},
	author = {Pinna, D. and Bourianoff, G. and Everschor-Sitte, K.},
	month = nov,
	year = {2020},
	pages = {054020},
}

@article{burkert_giant_2004,
	title = {Giant {Magnetic} {Anisotropy} in {Tetragonal} {FeCo} {Alloys}},
	volume = {93},
	url = {https://link.aps.org/doi/10.1103/PhysRevLett.93.027203},
	doi = {10.1103/PhysRevLett.93.027203},
	number = {2},
	urldate = {2026-06-22},
	journal = {Physical Review Letters},
	publisher = {American Physical Society},
	author = {Burkert, Till and Nordström, Lars and Eriksson, Olle and Heinonen, Olle},
	month = jul,
	year = {2004},
	pages = {027203},
}

@article{finocchio_magnetic_2016,
	title = {Magnetic skyrmions: from fundamental to applications},
	volume = {49},
	issn = {0022-3727},
	shorttitle = {Magnetic skyrmions},
	url = {https://doi.org/10.1088/0022-3727/49/42/423001},
	doi = {10.1088/0022-3727/49/42/423001},
	language = {en},
	number = {42},
	urldate = {2026-06-22},
	journal = {Journal of Physics D: Applied Physics},
	publisher = {IOP Publishing},
	author = {Finocchio, Giovanni and Büttner, Felix and Tomasello, Riccardo and Carpentieri, Mario and Kläui, Mathias},
	month = sep,
	year = {2016},
	pages = {423001},
}

@article{hrabec_measuring_2014,
	title = {Measuring and tailoring the {Dzyaloshinskii}-{Moriya} interaction in perpendicularly magnetized thin films},
	volume = {90},
	url = {https://link.aps.org/doi/10.1103/PhysRevB.90.020402},
	doi = {10.1103/PhysRevB.90.020402},
	number = {2},
	urldate = {2026-05-16},
	journal = {Physical Review B},
	publisher = {American Physical Society},
	author = {Hrabec, A. and Porter, N. A. and Wells, A. and Benitez, M. J. and Burnell, G. and McVitie, S. and McGrouther, D. and Moore, T. A. and Marrows, C. H.},
	month = jul,
	year = {2014},
	pages = {020402},
}

@article{kuepferling_measuring_2023,
	title = {Measuring interfacial {Dzyaloshinskii}-{Moriya} interaction in ultrathin magnetic films},
	volume = {95},
	url = {https://link.aps.org/doi/10.1103/RevModPhys.95.015003},
	doi = {10.1103/RevModPhys.95.015003},
	number = {1},
	urldate = {2026-06-22},
	journal = {Reviews of Modern Physics},
	publisher = {American Physical Society},
	author = {Kuepferling, M. and Casiraghi, A. and Soares, G. and Durin, G. and Garcia-Sanchez, F. and Chen, L. and Back, C. H. and Marrows, C. H. and Tacchi, S. and Carlotti, G.},
	month = mar,
	year = {2023},
	pages = {015003},
}

@article{greene_deposition_2014,
	title = {Deposition order dependent magnetization reversal in pressure graded {Co}/{Pd} films},
	volume = {104},
	issn = {0003-6951},
	url = {https://doi.org/10.1063/1.4871586},
	doi = {10.1063/1.4871586},
	number = {15},
	urldate = {2026-06-16},
	journal = {Applied Physics Letters},
	author = {Greene, P. K. and Kirby, B. J. and Lau, J. W. and Borchers, J. A. and Fitzsimmons, M. R. and Liu, Kai},
	month = apr,
	year = {2014},
	pages = {152401},
}

@article{kirby_vertically_2010,
	title = {Vertically graded anisotropy in {Co}/{Pd} multilayers},
	volume = {81},
	url = {https://link.aps.org/doi/10.1103/PhysRevB.81.100405},
	doi = {10.1103/PhysRevB.81.100405},
	number = {10},
	urldate = {2026-06-16},
	journal = {Physical Review B},
	publisher = {American Physical Society},
	author = {Kirby, B. J. and Davies, J. E. and Liu, Kai and Watson, S. M. and Zimanyi, G. T. and Shull, R. D. and Kienzle, P. A. and Borchers, J. A.},
	month = mar,
	year = {2010},
	pages = {100405},
}

@article{kirby_direct_2009,
	title = {Direct observation of magnetic gradient in {Co}/{Pd} pressure-graded media},
	volume = {105},
	issn = {0021-8979},
	url = {https://doi.org/10.1063/1.3077224},
	doi = {10.1063/1.3077224},
	number = {7},
	urldate = {2026-05-16},
	journal = {Journal of Applied Physics},
	author = {Kirby, B. J. and Watson, S. M. and Davies, J. E. and Zimanyi, G. T. and Liu, Kai and Shull, R. D. and Borchers, J. A.},
	month = mar,
	year = {2009},
	pages = {07C929},
}

@article{burks_3d_2021,
	title = {{3D} {Nanomagnetism} in {Low} {Density} {Interconnected} {Nanowire} {Networks}},
	volume = {21},
	issn = {1530-6984},
	url = {https://doi.org/10.1021/acs.nanolett.0c04366},
	doi = {10.1021/acs.nanolett.0c04366},
	number = {1},
	urldate = {2026-06-15},
	journal = {Nano Letters},
	publisher = {American Chemical Society},
	author = {Burks, Edward C. and Gilbert, Dustin A. and Murray, Peyton D. and Flores, Chad and Felter, Thomas E. and Charnvanichborikarn, Supakit and Kucheyev, Sergei O. and Colvin, Jeffrey D. and Yin, Gen and Liu, Kai},
	month = jan,
	year = {2021},
	pages = {716--722},
}

@article{gilbert_realization_2015,
	title = {Realization of ground-state artificial skyrmion lattices at room temperature},
	volume = {6},
	doi = {10.1038/ncomms9462},
	number = {1},
	journal = {Nature Communications},
	publisher = {Springer Science and Business Media LLC},
	author = {Gilbert, Dustin A. and Maranville, Brian B. and Balk, Andrew L. and Kirby, Brian J. and Fischer, Peter and Pierce, Daniel T. and Unguris, John and Borchers, Julie A. and Liu, Kai},
	month = oct,
	year = {2015},
	pages = {8462},
}

@article{rahman_controlling_2009,
	title = {Controlling magnetization reversal in {Co}/{Pt} nanostructures with perpendicular anisotropy},
	volume = {94},
	doi = {10.1063/1.3075061},
	number = {4},
	journal = {Applied Physics Letters},
	publisher = {AIP Publishing},
	author = {Rahman, M. Tofizur and Dumas, Randy K. and Eibagi, Nasim and Shams, Nazmun N. and Wu, Yun-Chung and Liu, Kai and Lai, Chih-Huang},
	month = jan,
	year = {2009},
	pages = {042507},
}

@article{dumas_magnetic_2007,
	title = {Magnetic fingerprints of sub-100 nm {Fe} dots},
	volume = {75},
	doi = {10.1103/physrevb.75.134405},
	number = {13},
	journal = {Physical Review B},
	publisher = {American Physical Society (APS)},
	author = {Dumas, Randy K. and Li, Chang-Peng and Roshchin, Igor V. and Schuller, Ivan K. and Liu, Kai},
	month = apr,
	year = {2007},
	pages = {134405},
}

@article{gilbert_quantitative_2014,
	title = {Quantitative {Decoding} of {Interactions} in {Tunable} {Nanomagnet} {Arrays} {Using} {First} {Order} {Reversal} {Curves}},
	volume = {4},
	doi = {10.1038/srep04204},
	number = {1},
	journal = {Scientific Reports},
	publisher = {Springer Science and Business Media LLC},
	author = {Gilbert, Dustin A. and Zimanyi, Gergely T. and Dumas, Randy K. and Winklhofer, Michael and Gomez, Alicia and Eibagi, Nasim and Vicent, J. L. and Liu, Kai},
	month = feb,
	year = {2014},
	pages = {4204},
}

@article{davies_magnetization_2004,
	title = {Magnetization reversal of {Co}/{Pt}: {Microscopic} origin of high-field magnetic irreversibility},
	volume = {70},
	doi = {doi.org/10.1103/PhysRevB.70.224434},
	number = {22},
	journal = {Physical Review B},
	publisher = {American Physical Society (APS)},
	author = {Davies, Joseph E. and Hellwig, Olav and Fullerton, Eric E. and Denbeaux, Greg and Kortright, J. B. and Liu, Kai},
	month = dec,
	year = {2004},
	pages = {224434},
}

@article{soucaille_probing_2016,
	title = {Probing the {Dzyaloshinskii}-{Moriya} interaction in {CoFeB} ultrathin films using domain wall creep and {Brillouin} light spectroscopy},
	volume = {94},
	url = {https://link.aps.org/doi/10.1103/PhysRevB.94.104431},
	doi = {10.1103/PhysRevB.94.104431},
	number = {10},
	urldate = {2026-05-16},
	journal = {Physical Review B},
	publisher = {American Physical Society},
	author = {Soucaille, R. and Belmeguenai, M. and Torrejon, J. and Kim, J.-V. and Devolder, T. and Roussigné, Y. and Chérif, S.-M. and Stashkevich, A. A. and Hayashi, M. and Adam, J.-P.},
	month = sep,
	year = {2016},
	pages = {104431},
}

@article{woo_observation_2016,
	title = {Observation of room-temperature magnetic skyrmions and their current-driven dynamics in ultrathin metallic ferromagnets},
	volume = {15},
	copyright = {2016 Springer Nature Limited},
	issn = {1476-4660},
	url = {https://www.nature.com/articles/nmat4593},
	doi = {10.1038/nmat4593},
	language = {en},
	number = {5},
	urldate = {2026-05-16},
	journal = {Nature Materials},
	publisher = {Nature Publishing Group},
	author = {Woo, Seonghoon and Litzius, Kai and Krüger, Benjamin and Im, Mi-Young and Caretta, Lucas and Richter, Kornel and Mann, Maxwell and Krone, Andrea and Reeve, Robert M. and Weigand, Markus and Agrawal, Parnika and Lemesh, Ivan and Mawass, Mohamad-Assaad and Fischer, Peter and Kläui, Mathias and Beach, Geoffrey S. D.},
	month = may,
	year = {2016},
	pages = {501--506},
}

@article{zhu_micromagnetic_1988,
	title = {Micromagnetic studies of thin metallic films (invited)},
	volume = {63},
	issn = {0021-8979},
	url = {https://doi.org/10.1063/1.341167},
	doi = {10.1063/1.341167},
	number = {8},
	urldate = {2026-05-16},
	journal = {Journal of Applied Physics},
	author = {Zhu, Jian‐Gang and Bertram, H. Neal},
	month = apr,
	year = {1988},
	pages = {3248--3253},
}

@article{kronmuller_micromagnetism_1996,
	title = {Micromagnetism and microstructure of hard magnetic materials},
	volume = {29},
	issn = {0022-3727},
	url = {https://doi.org/10.1088/0022-3727/29/9/008},
	doi = {10.1088/0022-3727/29/9/008},
	language = {en},
	number = {9},
	urldate = {2026-05-16},
	journal = {Journal of Physics D: Applied Physics},
	author = {Kronmüller, H. and Fischer, R. and Seeger, M. and Zern, A.},
	month = sep,
	year = {1996},
	pages = {2274},
}

@article{ferriani_atomic-scale_2008,
	title = {Atomic-{Scale} {Spin} {Spiral} with a {Unique} {Rotational} {Sense}: {Mn} {Monolayer} on {W}(001)},
	volume = {101},
	url = {https://doi.org/10.1103$\%$2Fphysrevlett.101.027201},
	doi = {10.1103/physrevlett.101.027201},
	number = {2},
	journal = {Physical Review Letters},
	publisher = {American Physical Society (APS)},
	author = {Ferriani, P. and Bergmann, K. von and Vedmedenko, E. Y. and Heinze, S. and Bode, M. and Heide, M. and Bihlmayer, G. and Blügel, S. and Wiesendanger, R.},
	month = jul,
	year = {2008},
	pages = {027201},
}

@article{nagaosa_topological_2013,
	title = {Topological properties and dynamics of magnetic skyrmions},
	volume = {8},
	copyright = {2013 Springer Nature Limited},
	issn = {1748-3395},
	url = {https://www.nature.com/articles/nnano.2013.243},
	doi = {10.1038/nnano.2013.243},
	language = {en},
	number = {12},
	urldate = {2026-05-16},
	journal = {Nature Nanotechnology},
	publisher = {Nature Publishing Group},
	author = {Nagaosa, Naoto and Tokura, Yoshinori},
	month = dec,
	year = {2013},
	pages = {899--911},
}

@article{abert_micromagnetics_2019,
	title = {Micromagnetics and spintronics: models and numerical methods},
	volume = {92},
	issn = {1434-6036},
	shorttitle = {Micromagnetics and spintronics},
	url = {https://doi.org/10.1140/epjb/e2019-90599-6},
	doi = {10.1140/epjb/e2019-90599-6},
	language = {en},
	number = {6},
	urldate = {2026-05-16},
	journal = {The European Physical Journal B},
	author = {Abert, Claas},
	month = jun,
	year = {2019},
	pages = {120},
}

@inproceedings{landau_theory_1992,
	title = {On the theory of the dispersion of magnetic permeability in ferromagnetic bodies},
	url = {https://www.sciencedirect.com/science/chapter/edited-volume/abs/pii/B9780080363646500089},
	doi = {10.1016/B978-0-08-036364-6.50008-9},
	language = {en-US},
	urldate = {2026-05-16},
	booktitle = {Perspectives in {Theoretical} {Physics}},
	publisher = {Pergamon},
	author = {Landau, Lev Davidovich and Lifshitz, Evgen Mikhailovichy},
	editor = {Pitaevskii, Lev Petrovich},
	month = jan,
	year = {1992},
	pages = {51--65},
}

@article{brown_micromagnetics_1959,
	title = {Micromagnetics, {Domains}, and {Resonance}},
	volume = {30},
	issn = {0021-8979},
	url = {https://doi.org/10.1063/1.2185970},
	doi = {10.1063/1.2185970},
	number = {4},
	urldate = {2026-05-16},
	journal = {Journal of Applied Physics},
	author = {Brown, Jr., William Fuller},
	month = apr,
	year = {1959},
	pages = {S62--S69},
}

@article{balasubramanian_synergistic_2020,
	title = {Synergistic computational and experimental discovery of novel magnetic materials},
	volume = {5},
	issn = {2058-9689},
	url = {https://pubs.rsc.org/en/content/articlelanding/2020/me/d0me00050g},
	doi = {10.1039/D0ME00050G},
	language = {en},
	number = {6},
	urldate = {2024-05-30},
	journal = {Molecular Systems Design \& Engineering},
	publisher = {The Royal Society of Chemistry},
	author = {Balasubramanian, Balamurugan and Sakurai, Masahiro and Wang, Cai-Zhuang and Xu, Xiaoshan and Ho, Kai-Ming and Chelikowsky, James R. and Sellmyer, David J.},
	month = jul,
	year = {2020},
	pages = {1098--1117},
}

@article{sakurai_discovering_2020,
	title = {Discovering rare-earth-free magnetic materials through the development of a database},
	volume = {4},
	url = {https://link.aps.org/doi/10.1103/PhysRevMaterials.4.114408},
	doi = {10.1103/PhysRevMaterials.4.114408},
	number = {11},
	urldate = {2024-05-30},
	journal = {Physical Review Materials},
	publisher = {American Physical Society},
	author = {Sakurai, Masahiro and Wang, Renhai and Liao, Timothy and Zhang, Chao and Sun, Huaijun and Sun, Yang and Wang, Haidi and Zhao, Xin and Wang, Songyou and Balasubramanian, Balamurugan and Xu, Xiaoshan and Sellmyer, David J. and Antropov, Vladimir and Zhang, Jianhua and Wang, Cai-Zhuang and Ho, Kai-Ming and Chelikowsky, James R.},
	month = nov,
	year = {2020},
	pages = {114408},
}

@article{jeudy_pinning_2018,
	title = {Pinning of domain walls in thin ferromagnetic films},
	volume = {98},
	url = {https://link.aps.org/doi/10.1103/PhysRevB.98.054406},
	doi = {10.1103/PhysRevB.98.054406},
	number = {5},
	urldate = {2024-05-13},
	journal = {Physical Review B},
	publisher = {American Physical Society},
	author = {Jeudy, V. and Díaz Pardo, R. and Savero Torres, W. and Bustingorry, S. and Kolton, A. B.},
	month = aug,
	year = {2018},
	pages = {054406},
}

@article{fugetta_machine-learning_2023,
	title = {Machine-learning recognition of {Dzyaloshinskii}-{Moriya} interaction from magnetometry},
	volume = {5},
	url = {https://link.aps.org/doi/10.1103/PhysRevResearch.5.043012},
	doi = {10.1103/PhysRevResearch.5.043012},
	number = {4},
	urldate = {2024-05-13},
	journal = {Physical Review Research},
	publisher = {American Physical Society},
	author = {Fugetta, Bradley J. and Chen, Zhijie and Bhattacharya, Dhritiman and Yue, Kun and Liu, Kai and Liu, Amy Y. and Yin, Gen},
	month = oct,
	year = {2023},
	pages = {043012},
}

@article{vansteenkiste_design_2014,
	title = {The design and verification of {MuMax3}},
	volume = {4},
	url = {https://doi.org/10.1063%2F1.4899186},
	doi = {10.1063/1.4899186},
	number = {10},
	journal = {AIP Advances},
	publisher = {AIP Publishing},
	author = {Vansteenkiste, Arne and Leliaert, Jonathan and Dvornik, Mykola and Helsen, Mathias and Garcia-Sanchez, Felipe and Waeyenberge, Bartel Van},
	month = oct,
	year = {2014},
	pages = {107133},
}

@article{hahn_temperature_2019,
	title = {Temperature in micromagnetism: cell size and scaling effects of the stochastic {Landau}–{Lifshitz} equation},
	volume = {3},
	issn = {2399-6528},
	shorttitle = {Temperature in micromagnetism},
	url = {https://dx.doi.org/10.1088/2399-6528/ab31e6},
	doi = {10.1088/2399-6528/ab31e6},
	language = {en},
	number = {7},
	urldate = {2023-07-31},
	journal = {Journal of Physics Communications},
	publisher = {IOP Publishing},
	author = {Hahn, Marc Benjamin},
	month = jul,
	year = {2019},
	pages = {075009},
}

@article{kwon_magnetic_2020,
	title = {Magnetic {Hamiltonian} parameter estimation using deep learning techniques},
	volume = {6},
	url = {https://doi.org/10.1126$\%$2Fsciadv.abb0872},
	doi = {10.1126/sciadv.abb0872},
	number = {39},
	journal = {Science Advances},
	publisher = {American Association for the Advancement of Science (AAAS)},
	author = {Kwon, H. Y. and Yoon, H. G. and Lee, C. and Chen, G. and Liu, K. and Schmid, A. K. and Wu, Y. Z. and Choi, J. W. and Won, C.},
	month = sep,
	year = {2020},
	pages = {eabb0872},
}

@article{wang_machine_2020,
	title = {Machine {Learning} {Magnetic} {Parameters} from {Spin} {Configurations}},
	volume = {7},
	url = {https://doi.org/10.1002%2Fadvs.202000566},
	doi = {10.1002/advs.202000566},
	number = {16},
	journal = {Advanced Science},
	publisher = {Wiley},
	author = {Wang, Dingchen and Wei, Songrui and Yuan, Anran and Tian, Fanghua and Cao, Kaiyan and Zhao, Qizhong and Zhang, Yin and Zhou, Chao and Song, Xiaoping and Xue, Dezhen and Yang, Sen},
	month = jul,
	year = {2020},
	pages = {2000566},
}

@book{mayergoyz_mathematical_1991,
	title = {Mathematical {Models} of {Hysteresis}},
	url = {https://doi.org/10.1007$\%$2F978-1-4612-3028-1},
	doi = {10.1007/978-1-4612-3028-1},
	publisher = {Springer New York},
	author = {Mayergoyz, I. D.},
	year = {1991},
}

@article{torrejon_interface_2014,
	title = {Interface control of the magnetic chirality in {CoFeB}/{MgO} heterostructures with heavy-metal underlayers},
	volume = {5},
	url = {https://doi.org/10.1038$\%$2Fncomms5655},
	doi = {10.1038/ncomms5655},
	number = {1},
	journal = {Nature Communications},
	publisher = {Springer Science and Business Media LLC},
	author = {Torrejon, Jacob and Kim, Junyeon and Sinha, Jaivardhan and Mitani, Seiji and Hayashi, Masamitsu and Yamanouchi, Michihiko and Ohno, Hideo},
	month = aug,
	year = {2014},
	pages = {4655},
}

@article{ma_dzyaloshinskii-moriya_2017,
	title = {Dzyaloshinskii-{Moriya} {Interaction} across an {Antiferromagnet}-{Ferromagnet} {Interface}},
	volume = {119},
	url = {https://doi.org/10.1103$\%$2Fphysrevlett.119.027202},
	doi = {10.1103/physrevlett.119.027202},
	number = {2},
	journal = {Physical Review Letters},
	publisher = {American Physical Society (APS)},
	author = {Ma, Xin and Yu, Guoqiang and Razavi, Seyed A. and Sasaki, Stephen S. and Li, Xiang and Hao, Kai and Tolbert, Sarah H. and Wang, Kang L. and Li, Xiaoqin},
	month = jul,
	year = {2017},
	pages = {027202},
}

@article{heide_dzyaloshinskii-moriya_2008,
	title = {Dzyaloshinskii-{Moriya} interaction accounting for the orientation of magnetic domains in ultrathin films: {Fe}/{W}(110)},
	volume = {78},
	url = {https://doi.org/10.1103$\%$2Fphysrevb.78.140403},
	doi = {10.1103/physrevb.78.140403},
	number = {14},
	journal = {Physical Review B},
	publisher = {American Physical Society (APS)},
	author = {Heide, M. and Bihlmayer, G. and Blügel, S.},
	month = oct,
	year = {2008},
	pages = {140403},
}

@article{kawaguchi_determination_2021,
	title = {Determination of the {Dzyaloshinskii}-{Moriya} interaction using pattern recognition and machine learning},
	volume = {7},
	url = {https://doi.org/10.1038$\%$2Fs41524-020-00485-2},
	doi = {10.1038/s41524-020-00485-2},
	number = {1},
	journal = {npj Computational Materials},
	publisher = {Springer Science and Business Media LLC},
	author = {Kawaguchi, Masashi and Tanabe, Kenji and Yamada, Keisuke and Sawa, Takuya and Hasegawa, Shun and Hayashi, Masamitsu and Nakatani, Yoshinobu},
	month = jan,
	year = {2021},
	pages = {20},
}

@article{bode_chiral_2007,
	title = {Chiral magnetic order at surfaces driven by inversion asymmetry},
	volume = {447},
	url = {https://doi.org/10.1038$\%$2Fnature05802},
	doi = {10.1038/nature05802},
	number = {7141},
	journal = {Nature},
	publisher = {Springer Science and Business Media LLC},
	author = {Bode, M. and Heide, M. and Bergmann, K. von and Ferriani, P. and Heinze, S. and Bihlmayer, G. and Kubetzka, A. and Pietzsch, O. and Blügel, S. and Wiesendanger, R.},
	month = may,
	year = {2007},
	pages = {190--193},
}

@article{pike_characterizing_1999,
	title = {Characterizing interactions in fine magnetic particle systems using first order reversal curves},
	volume = {85},
	url = {https://doi.org/10.1063$\%$2F1.370176},
	doi = {10.1063/1.370176},
	number = {9},
	journal = {Journal of Applied Physics},
	publisher = {AIP Publishing},
	author = {Pike, Christopher R. and Roberts, Andrew P. and Verosub, Kenneth L.},
	month = may,
	year = {1999},
	pages = {6660--6667},
}

@article{grinstein_coarse_2003,
	title = {Coarse {Graining} in {Micromagnetics}},
	volume = {90},
	url = {https://link.aps.org/doi/10.1103/PhysRevLett.90.207201},
	doi = {10.1103/PhysRevLett.90.207201},
	number = {20},
	urldate = {2023-07-31},
	journal = {Physical Review Letters},
	publisher = {American Physical Society},
	author = {Grinstein, G. and Koch, R. H.},
	month = may,
	year = {2003},
	pages = {207201},
}

@article{zakeri_asymmetric_2010,
	title = {Asymmetric {Spin}-{Wave} {Dispersion} on {Fe}(110): {Direct} {Evidence} of the {Dzyaloshinskii}-{Moriya} {Interaction}},
	volume = {104},
	url = {https://doi.org/10.1103$\%$2Fphysrevlett.104.137203},
	doi = {10.1103/physrevlett.104.137203},
	number = {13},
	journal = {Physical Review Letters},
	publisher = {American Physical Society (APS)},
	author = {Zakeri, Kh and Zhang, Y. and Prokop, J. and Chuang, T.-H. and Sakr, N. and Tang, W. X. and Kirschner, J.},
	month = mar,
	year = {2010},
	pages = {137203},
}

@misc{zeiler_adadelta_2012,
	title = {{ADADELTA}: {An} {Adaptive} {Learning} {Rate} {Method}},
	shorttitle = {{ADADELTA}},
	url = {http://arxiv.org/abs/1212.5701},
	doi = {10.48550/arXiv.1212.5701},
	urldate = {2023-07-31},
	publisher = {arXiv},
	author = {Zeiler, Matthew D.},
	month = dec,
	year = {2012},
	note = {arXiv:1212.5701 [cs]},
}

@article{singh_application_2019,
	title = {Application of machine learning to two-dimensional {Dzyaloshinskii}-{Moriya} ferromagnets},
	volume = {99},
	url = {https://doi.org/10.1103$\%$2Fphysrevb.99.174426},
	doi = {10.1103/physrevb.99.174426},
	number = {17},
	journal = {Physical Review B},
	publisher = {American Physical Society (APS)},
	author = {Singh, Vinit Kumar and Han, Jung Hoon},
	month = may,
	year = {2019},
	pages = {174426},
}
\end{document}